\documentclass[a4paper,11pt]{article}
\pdfoutput=1 
\usepackage{jheppub} 
\usepackage{amsmath}
\usepackage[T1]{fontenc} 
\usepackage{latexsym}
\usepackage{amssymb,amsfonts}
\usepackage{soul}
\usepackage{amsfonts}
\font\mybb=msbm10 at 11pt
\def\bb#1{\hbox{\mybb#1}}

\renewcommand{\theequation}{\arabic{section}.\arabic{equation}}

\preprint{D3nAmW-Ham-4.tex $\quad$
July 30, 2018  }
\title{Hamiltonian approach and quantization of $D=3, {\cal N}=1$ supersymmetric non-Abelian multiwave system}

\vspace{2.5cm}

\author[\dagger\ddagger]{Igor Bandos}
 \author[\dagger \ast]{and M. Sabido}
\affiliation[\dagger]{Department of Theoretical Physics, University of the Basque Country UPV/EHU,
\\
P.O. Box 644, 48080 Bilbao, Spain.}

\affiliation[\ddagger]{IKERBASQUE, Basque Foundation for Science, 48011, Bilbao, Spain.}
\affiliation[\ast]{ Department of Physics, University of Guanajuato, A.P. E-
143, C.P. 37150,
Leon, Guanajuato, Mexico. }

\date{}

\abstract{
We develop Hamiltonian formalism and quantize supersymmetric non-Abelian multiwave system (nAmW) in D=3 spacetime constructed as a simple counterpart of 11D multiple M-wave system. Its action can be obtained from massless superparticle one by putting on its worldline  1d  dimensional reduction of the 3d SYM model in such a way that the new system still possesses local fermionic kappa-symmetry.
\\
The quantization results in a set of equation of  supersymmetric field theory in an unusual space with $su(N)$--valued matrix coordinates. Their superpartners, the fermionic  $su(N)$--valued matrices, cannot be split  on coordinates and momenta in a covariant manner and hence are included as abstract operators acting on the state vector in the generic form of our D=3 Matrix model field equations. 
 We discuss the Clifford superfield representation for the quantum  state vector and in the simplest case of $N=2$ elaborate it in a bit more detail.
As a check of consistency, we show that the bosonic Matrix model field equations obtained  by quantization of the purely bosonic limit of our D=3 nAmW system have a nontrivial solution. 
 }


\begin{document}
\maketitle
\flushbottom

\section{Introduction}

The quantization of relativistic particle mechanics is known to result in free field theory. In the simplest case this is a scalar field
obeying free Klein--Gordon equations  (see e.g. sec. 1.1.1. of \cite{West:2012vka} or sec. 1.2. of \cite{Bandos:1993ma}). The mechanical model with worldline supersymmetry allows to obtain spinor fields obeying the Dirac equations \cite{Berezin:1976eg,Brink:1976sc,Brink:1976uf}\footnote{See \cite{Gershun:1979fb,Howe:1988ft} as well as \cite{Bastianelli:2005vk,Bastianelli:2005uy,Bastianelli:2007pv,Bonezzi:2018box} and refs. therein for derivation of equations for higher spin fields by the quantization of spinning particle model with extended worldline supersymmetry.
}. The quantization of supersymmetric particle mechanics, so--called superparticle models \cite{Brink:1981nb,de Azcarraga:1982dw,Siegel:1983hh}, results in a set of free equations for supermultiplets of fields. In higher supersymmetric and higher dimensional cases the supermultiplet without `higher spin' fields becomes unique. For instance, the light-cone gauge quantization of 11D superparticle mechanics results in the linearized 11D supergravity \cite{Green:1999by}  (see \cite{Bandos:2007mi,Bandos:2007wm} for covariant quantization).

The superparticle is the simplest ($p$=0) representative of the family of supersymmetric $p$--branes or super-$p$-branes. Higher ($p>0$) $p$-branes are supersymmetric extended objects with $(p+1)$ dimensional worldvolume. An especial role in the set of 10-dimensional branes {is} played by fundamental superstring
(F$1$-brane) and Dirichlet $p$-branes (D$p$-branes) on  which  the fundamental string can have its endpoints. As we were {taught} by Witten
\cite{Witten:1995im}, the system of multiple (almost) coincident D$p$-branes carries non-abelian gauge fields and at very low energy, is described by a $U(N)$ supersymmetric gauge theory. The problem of finding a complete action for multiple D$p$-brane system is still open, although some progress was reached during these years  \cite{Tseytlin:1997csa,Myers:1999ps,Lozano:2005kf,YLozano+=0207,Howe+Lindstrom+Linus}. {The commonly accepted bosonic limit  is given by the  Myers's `dielectric brane' action}  \cite{Myers:1999ps}.

A complete action including fermions and invariant under spacetime supersymmetry and local fermionic $\kappa$-symmetry (which guaranties that the ground state of the system is a BPS state) is known for the system of $D\leq 10$ multiple $0$-branes
\cite{Sorokin:2001av,Panda:2003dj} {as well as} for 11D multiple M0-brane (multiple M-waves  or mM0) system
\cite{Bandos:2012jz,Bandos:2013uoa}. The dimensional reduction of this has to reproduce a multiple D$0$-brane  (mD$0$) action\footnote{Notice also a very interesting construction on '-1 quantization level' \cite{Howe+Lindstrom+Linus}; this is to say the quantization of the dynamical system of \cite{Howe+Lindstrom+Linus} should reproduce multiple D$p$-brane action.}. Furthermore, in {the} three dimensional case, the complete supersymmetric  non-Abelian Born-Infeld action \cite{Drummond:2002kg}, {and} non-Abelian multiwave action \cite{Bandos:2013swa}, the 3d counterpart of the mM0 {action, are known}\footnote{Notice also the existence of  Bagger--Lambert--Gustavsson (BLG) model
\cite{Bagger:2006sk,Gustavsson:2007vu} and Aharony--Bergman--Jafferis--Maldacena (ABJM) model \cite{Aharony:2008ug}, both associated to the infrared fixed points of multiple M2-brane systems. Also an interesting conjecture of that the $D=6$ $(2,0)$ superconformal theory, associated with multiple M$5$-brane system, can be described by
$D=5$ supersymmetric Yang-Mills (SYM) model was proposed in  \cite{Douglas:2010iu,Lambert:2010iw}.}.

The dynamical system of \cite{Bandos:2012jz,Bandos:2013uoa} can be considered as 11D massless superparticle carrying on its worldvolume
a 1d multiplet obtained by dimensional reduction of the maximal 10D SU(N) SYM multiplet. Thus it can be considered as reparametrization invariant and locally supersymemtric (with respect to worldline supersymmetry) generalization of the Matrix model of \cite{deWit:1988wri,Banks:1996vh}.
Then an interesting question arises: what field theory will appear as a result of its quantization? This would be the free field theory of the 11D Matrix model.

As {previously mentioned}, the quantization of the massless  11d superparticle, which can be called M-wave, results in linearized 11D supergravity.
The wave function of the multiple M-wave  of \cite{Bandos:2012jz,Bandos:2013uoa} should depend on additional coordinates which suggests that besides 11d supergravity, some additional `higher spin' multiplet might appear. Such a study might produce an insight about some specific  degrees of freedom of the hypothetical underlying M-theory.

In this paper we solve a simpler counterpart of the above problem. We quantize a 3d counterpart of the 11D multiple M-waves, a so-called   non-Abelian multiwave system (nAmW) whose action was found in \cite{Bandos:2013swa}. Such a study can be considered as toy model for the quantization of multiple M-wave system, which we are planning to address in the future, but is also interesting on its own.

The quantization of the model results in an unconventional  free field theory on a (super)spacetime enlarged by a set of $su(N)$ valued matrix coordinates. We will try to restrain ourselves from the use of the word `non-commutative spacetime' as this is attributed to  the well known more complicated constructions of  \cite{Snyder:1946qz,Connes:1994yd,Connes:1997cr,Seiberg:1999vs} (also referred to as  ``non-commutative geometry'').
In our simple model only one bosonic and one fermionic matrix coordinates are present. In the Hamiltonian formalism the first is supplied by one  matrix momentum while the second plays the role of the momentum for itself and, thus, enters the generic form of the equations as an abstract operator obeying 
the Clifford algebra\footnote{Curiousely enough, an alternative of superparticle model with Clifford algebra valued fermionic fields was considered  in 1980 by Brink and Schwarz \cite{Brink:1980nz}.}. We consider briefly Clifford superfield realization of this algebra, in which the state vector of the superparticle model depends on the Clifford algebra valued variables and in the simplest case of N=2 we elaborate this description in more detail. We notice that in higher dimensional cases, beginning $D=4$ nAmW model (still to be constructed) an even number of fermionic matrix coordinate operators will appear so that a more economic holomorphic representation, with state vector depending on a half of fermionic matrix coordinate (anticommuting among themselves) will be  available.

We also studied the system of bosonic equations which can be obtained by  quantization  of the bosonic limit of our D=3 nAmW model.
Particularly, we show that  these  equations have a nontrivial solution.

\section{The action for D=3 nAmW and its supersymmetry}

In this paper we study the Hamiltonian mechanics and quantization of the  non-Abelian multiwave (nAmW) system  in flat D=3 superspace described in \cite{Bandos:2013swa}.

The action of this system can be written in the following from
\begin{eqnarray}
\label{SmM0=3D} S_{nAmW} &=& \int\limits_{W^1} \left(
\rho^{\#}{E}^{=} + {E}^{\#} {\cal H} {1\over \rho^{\#}}+
tr\left(- {\bb P}{D} {\bb X} + {i\over 4} { \Psi} {D} {\Psi}  \right) +   {i\mu \over \sqrt{\rho^{\#}}} {E}^{+} {\cal S} + {{\cal D} \rho^{\#} \over \rho^{\#}} {\cal J} \right). \;
\qquad
\end{eqnarray}
In it the integration is performed over a worldline $W^1$ parametrized by a proper time coordinate $\tau$ and defined as a line in flat superspace with the use of ``center of energy'' coordinate functions: bosonic vector  $x^\mu=x^\mu(\tau)$ and fermionic spinor $\theta^\alpha=\theta^\alpha(\tau)$. These are included
in ${E}^{=}$, ${E}^{\#} $ and ${E}^{+}$ as we describe below.

The action contains also  $N\times N$ traceless matrix fields:
bosonic $ {\bb P}(\tau)$ and  $ {\bb X}(\tau)$  and fermionic $\Psi (\tau)$ \footnote{These are related to the fields  used in \cite{Bandos:2013swa}, which are of different dimension and carry non vanishing $SO(1,1)$ weights,  by  $ {\bb X}= {\rho^{\#}\over \mu^2} {\bb X}_\#$,  $ {\bb P}=  {(\rho^{\#})^2\over \mu^4}  {\bb P}_{\#\#}$ and
$ {\Psi}=  {(\rho^{\#})^{3/2}\over \mu^3}  {\Psi}_{\#}^{-}$. Also the ``relative motion Hamiltonian'' ${\cal H}$ in  (\ref{HSYM=3D}) is related to the expression denoted by the same symbol in \cite{Bandos:2013swa} by ${\cal H}= {(\rho^{\#})^4\over \mu^6}  {\cal H}_{\#\#\#\#}$}. These together with
the 1d $SU(N)$ gauge potential ${\bb A}=d\tau {\bb A}_\tau (\tau) $ can be obtained from dimensional reduction of 3d supersymmetric gauge theory.
The   imaginary  traceless  $N\times N$ matrix gauge field ${\bb A}_\tau $ enters  covariant derivatives of the bosonic and fermionic matrix fields in the third term of the action,
\begin{equation}
\label{DbbX=}
D{\bb X}=d{\bb X} + [{\bb A}, {\bb X}]\; , \qquad  D{\Psi}=d{\Psi} + [{\bb A}, {\Psi}]\;
 \; .  \qquad
\end{equation}
The matrix fields  are also used to construct the  bosonic and fermionic currents
\begin{eqnarray}
\label{HSYM=3D} && {\cal H}=   {\mu^2 \over 2} tr\left( {\bb P} {\bb P} \right)  - 2\, \mu^2  tr\left({\bb X}\, \Psi{\Psi}\right), \;
\qquad
  \end{eqnarray}
and
\begin{eqnarray}
\label{cJ=} && {\cal S}=   tr\left( {\bb P} \Psi \right)\; , \qquad {\cal J}=   tr\left( {\bb P} {\bb X} \right)\; . \qquad
   \end{eqnarray}
The bosonic current (\ref{HSYM=3D}) contains a  dimensional  constant $\mu$, $[\mu ]=L^{-1}$, which also enters the fourth term of the action (\ref{SmM0=3D}).

The action also  contains vector and spinor moving frame fields which
are included, first of all, in the 1--forms
\begin{eqnarray}\label{E==Eu=}
{E}^{=}:= {E}^a u_{ {a}}^{=}\; ,  \qquad \label{E-=Ev-}
{E}^{-}:= {E}^\alpha v_{ {\alpha}}^{-}\; ,  \qquad
\\ \label{E++=Eu++} {E}^{\#}:= {E}^a u_{ {a}}^{\#}\; ,  \qquad
  {E}^{+}:= {E}^\alpha v_{ {\alpha}}^{+},\;   \qquad
\end{eqnarray}
three of which enter the first, second and fourth terms of the action.
These 1-forms are contractions of the  pull--backs of the bosonic and fermionic supervielbein forms of flat $D=3$ ${\cal N}=1$  superspace
\begin{eqnarray}
\label{hEal=3D}{E}^a =d{x}^a- id{\theta} \gamma^a {\theta}\; , \qquad {E}^\alpha =d{\theta}^\alpha\; ,
\qquad
\end{eqnarray}
constructed form the bosonic and fermionic coordinate functions ${x}^a={x}^a(\tau)$ and ${\theta}^\alpha={\theta}^\alpha(\tau)$,
with  (anti-respectively)  bosonic spinors $v_{ {\alpha}}^{\mp}$ and composite light-like vectors
\begin{eqnarray}
\label{u--=3D} u_a^{=}= v^-\gamma_av^- \qquad \Leftrightarrow \qquad 2 v^-_\alpha v^-_\beta = u_a^{=} \gamma^a_{\alpha\beta} \;
, \qquad u^{a=} u_a^{=}=0\; , \qquad
\\
\label{u++=3D} u_a^{\#}= v^+\gamma_av^+ \qquad \Leftrightarrow \qquad 2 v^+_\alpha v^+_\beta = u_a^{\#}
\gamma^a_{\alpha\beta} \; , \qquad u^{a\#} u_a^{\#}=0\; . \qquad
\end{eqnarray}
The bosonic spinors $v_{ {\alpha}}^{\mp}$ are normalized by
\begin{eqnarray}\label{v-v+=1}
v^{\alpha -} v_{ {\alpha}}^{+}:=\epsilon^{\alpha\beta} v_{ {\alpha}}^{+}v_{ {\beta}}^{-} =1\; .  \qquad
\end{eqnarray}
This allows us to identify them with columns of unimodular 2$\times$2 $SL(2,{\bb R})$-valued matrix
\begin{eqnarray}\label{Vharm-in-}
V_\alpha^{(\beta)}=  ( v_{ {\alpha}}^{+},  v_{ {\alpha}}^{-})\quad \in  \quad SL(2,{\bb R}), \;
\end{eqnarray}
and to call them D=3 spinor moving frame variables or Lorentz harmonics (see \cite{Bandos:2013swa} and references therein for more details and discussion).

Eqs. (\ref{v-v+=1}) implies that the  composite null vectors (\ref{u--=3D}) and (\ref{u++=3D}) are normalized
\begin{eqnarray}\label{u--u++=2}
u^{a\#} u_{a}^{=}=2\; ,    \qquad
\end{eqnarray}
which allow to identify them with elements of vector frame matrix
\begin{eqnarray}
\label{U=harm}
 U_a^{(b)}=\left( \frac 1 2 \left(u_a^{\#} +u_a^{=} \right),  u_a^{\perp}, \frac 1 2 \left(u_a^{\#} -u_a^{=} \right) \right)\quad \in \quad SO(1,2)\; .
  \qquad
\end{eqnarray}
The third vector $u_a^{\perp}$ in (\ref{U=harm}) is spacelike, normalized on minus unity, orthogonal to the light-like vectors of the frame, and  also composed from spinor moving frame variables:
\begin{eqnarray}
\label{up=3D} u_a^{\perp}= v^-\gamma_av^+ \qquad \Leftrightarrow \qquad 2 v^-_{(\alpha} v^+_{\beta )} = u_a^{\perp} \gamma^a_{\alpha\beta} \;
, \qquad
\\
\label{upup=1}  u^{a\perp} u_a^{\perp}=-1\; , \qquad u^{a\#} u_a^{\perp}=0\; , \qquad  u^{a=} u_a^{\perp}=0\; .  \qquad
\end{eqnarray}
Eq. (\ref{U=harm}) implies $ U_a^{(c)}  \eta^{ab} U_b^{(d)}=  \eta^{(c)(d)} $ which is equivalent to (\ref{upup=1}), (\ref{u--u++=2}) and the light-likeness conditions in (\ref{u--=3D}) and (\ref{u++=3D}).

The expressions for the light-like and spacelike vectors in (\ref{u--=3D}), (\ref{u++=3D}) and (\ref{up=3D}) can be obtained form the conditions of the $SO(1,2)$ Lorentz invariance of  3d gamma matrices,
$ U_a^{(b)}\tilde{\gamma}_{(b)}= V^T \tilde{\gamma}_a V$ and $ U_a^{(b)}{\gamma}^{a}= V{\gamma}^{(b)} V^T$.

Finally, four terms of the action contain scalar density  $\rho^{\#}=\rho^{\#}(\tau)$ which  is a Sht\"{u}ckelberg field for an $SO(1,1)$ gauge symmetry the weights of which are represented by sign indices. Its covariant derivative  includes the composite $SO(1,1)$ connection $\omega=  v^{\alpha -}d v_\alpha^+ =  v^{\alpha +}d v_\alpha^-$ the meaning of which will be clarified below in sec. \ref{smf-sec},
\begin{eqnarray}
\label{cDrho=}
{\cal D}\rho^{\#}= d\rho^{\#} - 2 {\omega}\rho^{\#}\; , \qquad \omega=  v^{\alpha -}d v_\alpha^+\; . \qquad
\end{eqnarray}

By construction, the  action (\ref{SmM0=3D}) is invariant under spacetime supersymmetry acting on the 'center of energy' coordinates 
only, $\delta_\varepsilon x^a = i\theta\gamma^a\varepsilon$, $\delta_\varepsilon \theta^\alpha=\varepsilon^\alpha$.  

The main property of the action (\ref{SmM0=3D}) is its invariance under the following local supersymmetry (SUSY) transformations, with the dimensionless fermionic parameter $\epsilon =\epsilon (\tau)$ \cite{Bandos:2013swa}
\begin{eqnarray}
\label{susy-X}  \delta_\epsilon {\bb X}   &=& i \epsilon \Psi \; , \quad \delta_\epsilon {\bb P}   = 0\; ,\qquad
\label{susy-Psi}  \delta_\epsilon \Psi =  2\epsilon {\bb P}\; ,\qquad \\
\label{susy-A}
 && \delta_\epsilon {\bb A} = -  4{E}^{\#} \epsilon \; {\mu^2 \over \rho^{\#}}  \Psi + 16 {E}^{+}
 \epsilon\; {\mu \over \rho^{\#}\sqrt{\rho^{\#}}}      {\bb X}
 \; ,  \qquad \\ \label{susy-rho}  &&  \delta_\epsilon \rho^{\#} = 0\; , \qquad \label{susy-v}
 \delta_\epsilon  v_{\alpha}^{\pm} =0
  \quad  \Rightarrow  \quad  \delta_\epsilon
  u_a^{=}= \delta_\epsilon u_a^{\#}= \delta_\epsilon u_a^{\perp}=0\;
 ,  \qquad
 \\ \label{susy-Z}
 \delta_\kappa {Z}^M &= & \epsilon \;  {\sqrt{\rho^{\#}}\over \mu } v^{-\alpha} E_{\alpha}^M ({Z})  +  {3i\over 4 \rho^{\#}}  \,
\epsilon \; tr(\Psi {\bb P}  )\;   u^{a\#}  E_{a}^M ({Z}) \; . \qquad
\end{eqnarray}
If we redefine the supersymmetry parameter by introducing $\kappa^+= {\sqrt{\rho^{\#}}\over \mu } \epsilon$ of dimension $[\kappa^+]= L^{1/2}$,  the transformations of the center of energy variables read
\begin{eqnarray}
\label{susy-th}
\delta_\kappa {Z}^M &= & \kappa^+  v^{-\alpha} E_{\alpha}^M ({Z})  +  {3i\mu \over 4\rho^{\#}\sqrt{\rho^{\#}} }  \,
\kappa^+ \; tr(\Psi {\bb P}  )\;    u^{a\#}  E_{a}^M ({Z})  \; , \qquad \kappa^+= {\sqrt{\rho^{\#}}\over \mu } \epsilon \; . \qquad
\end{eqnarray}
In this form it is clear that setting $N=1$, which implies the absence of matrix variables  ${\bb X}$,
${\bb P}$ and $\Psi$, we arrive at the characteristic expression  for the $\kappa$ symmetry of the massless superparticle in its twistor like spinor moving frame formulation.  Although this is related to the D=3 version of the Ferber-Schirafuji model \cite{Ferber:1977qx,Shirafuji:1983zd} by a field redefinition, it is more convenient as a basis for constructing the non-Abelian multiwave model (see \cite{Bandos:2013swa} for comparison).

In this paper we will develop the generalized Hamiltonian formalism for the above  nAmW system and perform its quantization.
To this end a few more details on spinor moving frame variables (\ref{Vharm-in-})
will be useful.


\subsection{More on spinor moving frame variables: covariant derivatives and Cartan forms}
\label{smf-sec}

In this subsection we describe some notions which were necessary to develop a convenient method of working with bosonic spinors constrained by  (\ref{v-v+=1}). Here they will allow us to lighten the discussion of  the Hamiltonian formalism.

The covariant derivatives that leave invariant the  normalization (\ref{v-v+=1}) are ({\it cf.} \cite{Galperin:2001uw} and \cite{B90})
\begin{eqnarray}\label{bbD++}
{\bb D}^{=}= v_\alpha^- {\partial\over \partial v_\alpha^+ }\; ,  \qquad {\bb D}^{\#}= v_\alpha^+ {\partial\over \partial v_\alpha^- }\; ,  \qquad {\bb D}^{(0)}= v_\alpha^+ {\partial\over \partial v_\alpha^+ }- v_\alpha^- {\partial\over \partial v_\alpha^- }\; .  \qquad
\end{eqnarray}
It is easy to check that the commutator of the covariant derivatives  form the $sl(2,{\bb R})$ algebra.
We can also see that
\begin{eqnarray}\label{bbDv=}
{\bb D}^{=} v_\alpha^+= v_\alpha^- \; &,& \qquad  {\bb D}^{=} v_\alpha^-= 0 \; , \qquad  \nonumber \\
{\bb D}^{\#} v_\alpha^+= 0 \; &,& \qquad  {\bb D}^{\#} v_\alpha^-= v_\alpha^+  \; , \qquad   \nonumber \\
&&  {\bb D}^{(0)} v_\alpha^\pm= \pm v_\alpha^\pm  \; . \qquad
\end{eqnarray}

As the covariant derivatives (\ref{bbD++})  preserve the constraint (\ref{v-v+=1}), using them as the only allowed differential operator on the $v_\alpha^\pm$ space, we can consider the functions of $v_\alpha^\pm$ obeying this constraints. Similarly, when constructing the generalized Hamiltonian mechanics of our system, in which (\ref{v-v+=1}) will be one of the constraints, we will
introduce the classical counterparts of (\ref{bbD++}), which can be called covariant momenta. When we formulate our mechanics in terms of such a covariant momenta, we can treat the constraint (\ref{v-v+=1})  as strong relations in the sense of Dirac \cite{Dirac:1963}.

The covariant derivatives  (\ref{bbD++}) can be obtained as coefficients of a decomposition of the differential in the space of
$v_{ {\alpha}}^{\pm}$ variables over the $SL(2,{\bb R})$ Cartan forms
\begin{eqnarray} \label{f++=}
 && f^{\#}= v^{\alpha +}d v_\alpha^+ \; , \qquad \\ \label{f==}
 && f^{=}= v^{\alpha -}d v_\alpha^- \; , \qquad \\ \label{om0=}
  && \omega = v^{\alpha -}d v_\alpha^+ =  v^{\alpha +}d v_\alpha^-\; ,   \qquad
\end{eqnarray}
namely
\begin{eqnarray}\label{d=Om--D+++}
d_{_{v^{\pm}}}&:=& dv_\alpha^\pm \frac \partial {\partial v_\alpha^\pm }
= -f^{\#}{\bb D}^{=}+ f^{=}{\bb D}^{\#} +  \omega {\bb D}^{(0)}\; . \qquad\end{eqnarray}

To check (\ref{d=Om--D+++}) one needs just to rewrite $dv_\alpha^{\pm}$ with the use of the identity
 \begin{eqnarray}\label{I=}
\delta_\alpha{}^\beta = v_\alpha^+ v^{\beta -} -v_\alpha^- v^{\beta +} \; ,  \qquad
\end{eqnarray}
which follows from (\ref{v-v+=1}),
 \begin{eqnarray}\label{dv=}
 dv_\alpha^+ = \omega v_\alpha^+ - f^{\#} v_\alpha^- \; ,  \qquad dv_\alpha^- = -\omega v_\alpha^- + f^{=} v_\alpha^+ \; .  \qquad
\end{eqnarray}
These  equations  can be also obtained from (\ref{d=Om--D+++}) and (\ref{bbDv=}).

\section{Hamiltonian formalism for nAmWs}

 The purpose of this section is to construct the generalized  Hamiltonian approach for the nAmW system.

\subsection{Momenta, Poisson brackets and canonical Hamiltonian }

The canonical momenta are defined as derivative of the Lagrangian over velocity. In particular, in our case the  momenta conjugate to the  bosonic and fermionic `center of energy' coordinate functions are defined by
\begin{eqnarray}
\label{Pa=dL-ddx}
P_a= {\partial {\cal L}\over \partial  \dot{x}{}^a}\; , \qquad
\pi_\alpha ={\partial {\cal L}\over \partial  \dot{ {\theta}}{}^\alpha}\; ,
\qquad
\end{eqnarray}
where $\dot{x}{}^a= \frac d {d\tau} {x}{}^a$, $\dot{ {\theta}}{}^\alpha=\frac d {d\tau}{ {\theta}}{}^\alpha$ and
\begin{eqnarray}
\label{LmM0=3D} {\cal L}:=
\rho^{\#}{E}_\tau^{=} + {E}_\tau^{\#} {\cal H} {1\over \rho^{\#}}+
tr\left(- {\bb P}{\cal D}_\tau {\bb X} + {i\over 4} { \Psi} {\cal D}_\tau {\Psi}  \right) +   {i\mu \over \sqrt{\rho^{\#}}} {E}_\tau^{+} {\cal S} + {{\cal D}_\tau \rho^{\#} \over \rho^{\#}} {\cal J} \; ,
\qquad
  \end{eqnarray}
is the Lagrangian of the nAmW action (\ref{SmM0=3D}),  $S_{nAmW}=\int d\tau  {\cal L}$.
The canonical Hamiltonian $H_0$ is given by the Legendre transformation of this Lagrangian,
\begin{eqnarray}
\label{H=Pdx-L}
H_0 =   \dot{x}^a P_a +  \dot{{\theta}}{}^\alpha \pi_\alpha + ... -  {\cal L}\; , \qquad
\qquad
\end{eqnarray}
where $...$ denotes the $\dot{q}p$ contributions from the other degrees of freedom which we are going to discuss below.

With the canonical definition of the Poisson brackets,
\begin{eqnarray}
\label{PB=Px}
{} [P_a, x^b]_{_{PB}}= - \delta_a{}^b\; , \qquad \{ \pi_\alpha \, , {\theta}^\beta\}_{_{PB}}=  - \delta_\alpha  {}^\beta\; ,
\qquad
\end{eqnarray}
the Hamiltonian equations of motion are
\begin{eqnarray}
\label{Eqs=Ham}
{} \dot{x}^a= [x^a, H]_{_{PB}} \, , \quad  \dot{\theta}^\alpha= [{\theta}^\alpha, H]_{_{PB}} \, , \quad  \dot{P}_a= [P_a, H]_{_{PB}}\; , \quad \dot{\pi}_\alpha= [{\pi}_\alpha, H]_{_{PB}} \; , \quad  ...
\quad
\end{eqnarray}

For the spinor moving frame variables it is convenient to use the $\tau$ components of the Cartan forms (\ref{f++=})--(\ref{om0=}),
$f^\#_\tau = v^{\alpha +}\partial_\tau v_\alpha^+$, $f^=_\tau = v^{\alpha -}\partial_\tau v_\alpha^-$ and
$\omega_\tau =v^{\alpha -}\partial_\tau v_\alpha^+$,   instead of velocities. Indeed, if we write the canonical expression and use the consequence (\ref{I=}) of the constraint (\ref{v-v+=1}), we find
\begin{eqnarray}
\label{vPv=fbfd}
\dot{v}_\alpha^+ P_{v^+}^\alpha + \dot{v}_\alpha^- P_{v^-}^\alpha = f_\tau ^\# {\mathbf d}^{=}- f_\tau^= {\mathbf d}^{\#} - \omega_\tau {\mathbf d}^{(0)} , \qquad
\end{eqnarray}
where the covariant momenta
\begin{eqnarray}
\label{bfd=dL-df}
{\mathbf d}^{=}= {\partial {\cal L}\over \partial f_\tau ^\#}  \; , \qquad {\mathbf d}^{\#} = {\partial  {\cal L} \over \partial f_\tau ^=} \; , \qquad {\mathbf d}^{(0)} = {\partial  {\cal L}\over \partial  \omega_\tau} \; , \qquad
\end{eqnarray}
are constructed from canonical ones as
\begin{eqnarray}
\label{d--:=}
{\mathbf d}^{=}= - {v}_\alpha^- P_{v^+}^\alpha, \qquad {\mathbf d}^{\#}= - {v}_\alpha^+ P_{v^-}^\alpha\quad and
\quad
{\mathbf d}^{(0)}= - ({v}_\alpha^+ P_{v^+}^\alpha - {v}_\alpha^- P_{v^-}^\alpha).
\end{eqnarray} One can easily check that
 the canonical Poisson brackets imply
\begin{eqnarray}
\label{bfdv=}
[ {\mathbf d}^{=}, {v}_\alpha^+ ]_{_{PB}}= {v}_\alpha^- , \qquad [ {\mathbf d}^{\#}, {v}_\alpha^- ]_{_{PB}}= {v}_\alpha^+, \qquad [ {\mathbf d}^{(0)}, {v}_\alpha^\pm ]_{_{PB}}= \pm {v}_\alpha^\pm \; , \qquad \\ \nonumber  [ {\mathbf d}^{\#}, {v}_\alpha^+ ]_{_{PB}}=0=[ {\mathbf d}^{=},{v}_\alpha^- ]_{_{PB}}\; , \qquad
\end{eqnarray}
or,
\begin{eqnarray}
\label{bfdv=}
[ {\mathbf d}^{=}, g(v^{\pm}) ]_{_{PB}}=  {\bb D}^{=} g(v^{\pm}) , \qquad [ {\mathbf d}^{\#}, g(v^{\pm}) ]_{_{PB}}=  {\bb D}^{\#} g(v^{\pm}) , \qquad \nonumber \\
{} [ {\mathbf d}^{(0)}, g(v^{\pm}) ]_{_{PB}}=  {\bb D}^{(0)} g(v^{\pm})\; , \qquad
\end{eqnarray}
where $ g(v^{\pm})$ is an arbitrary functions of spinor moving frame variables (i.e. of the bosonic spinors $v^\pm_\alpha$ obeying (\ref{v-v+=1})) and  $ {\bb D}^{=}, {\bb D}^{\#}$  and $ {\bb D}^{(0)}$ are covariant derivatives (\ref{bbD++}).

{It is straightforward to verify} that the covariant momenta have vanishing Poisson brackets with the constraint (\ref{v-v+=1}) so that, using only these (and not the forth independent linear combination
 $K= {v}_\alpha^+ P_{v^+}^\alpha + {v}_\alpha^- P_{v^-}^\alpha$
 of the momenta $P^\alpha_{v^\pm}$) we can consider this constraint as valid `in the strong sense'  \cite{Dirac:1963}, {\it i.e.} we can  impose it before calculating Poisson brackets \footnote{It is not hard to notice that this approach is equivalent to passing to the Dirac brackets with respect to the pair of second class constraints $\Xi= v^{-\alpha}v^{+}_{\alpha}-1\approx 0$ and $K= {v}_\alpha^+ P_{v^+}^\alpha + {v}_\alpha^- P_{v^-}^\alpha\approx 0$ (see \cite{Bandos:1993ma}). }.

 The canonical momentum for the bosonic matrix field ${\bb X}= {\bb X}_i{}^j $ can be defined as
\begin{eqnarray}
\label{PX=}
 {\bb P}_j{}^i =  - {\partial  {\cal L}\over \partial  \dot{{\bb X}}{}_i{}^j}.
\qquad
\end{eqnarray}
With this sign it can be identified with the traceless matrix ${\bb P}_j{}^i$ already included in the action (\ref{SmM0=3D}) so that we just state that the Poisson brackets between these two variables are given by
   \begin{eqnarray}
\label{PXX=PB}
 {}[ {\bb P}_i{}^j,  {\bb X}{}_{i'}{}^{j'}]_{_{PB}} = +\delta _i{}^{j'}\delta_{i'}{}^j -
 {1\over N}
 \delta_i{}^j\delta_{i'}{}^{j'} \; .
\qquad
\end{eqnarray}
To lighten the equations we can introduce `reference matrices', {these will be} traceless $N\times N$ bosonic matrices   ${\bb Y}= {\bb Y}_i{}^j $ and  ${\bb Z}= {\bb Z}_i{}^j $, and encode (\ref{PXX=PB}) in the relation
  \begin{eqnarray}
\label{YPXZ=PB}
 {}[ tr ({\bb Y}{\bb P}),  tr({\bb X} {\bb Z})]_{_{PB}} = tr ( {\bb Y}{\bb Z}) \;
\qquad
\end{eqnarray}
(the second term  in the {\it r.h.s.} of (\ref{PXX=PB}) does not contribute due to  $tr{\bb Y}=0=tr{\bb Z}$).

The canonical momentum for the fermionic matrix variable $\Psi=\Psi_i{}^j$ coincides with
these variables up to a numerical {factor}, $(\Pi_\Psi )_i{}^j = {\partial  {\cal L}\over \partial  \dot{{\Psi }}{}_j{}^i}=-{i\over 4}\Psi_i{}^j$. As usual, this  is reflected by stating that the fermionic variables have nonvanishing Poisson brackets {among themselves}\footnote{Actually, this implies passing to Dirac brackets with respect to the self-conjugate fermionic matrix constraint $(\Pi_\Psi + {i\over 4}\Psi)_i{}^j\approx 0$:
 $[...,...\}_{_{DB}}= [...,...\}_{_{PB}} - 2i [..., (\Pi_\Psi + {i\over 4}\Psi)_i{}^j \}_{_{PB}}[ (\Pi_\Psi + {i\over 4}\Psi)_j{}^i,...\}_{_{PB}}$.
  }. In our case these Poisson brackets are
\begin{eqnarray}
\label{YPsiZPsi=}
 {}\{ tr ({\bb Y}{\Psi}),  tr({\Psi} {\bb Z})\}_{_{PB}} = -2i tr ( {\bb Y}{\bb Z}). \;
\qquad
\end{eqnarray}

{Recapitulating}, the canonical Hamiltonian of our system can be defined by
\begin{eqnarray}
\label{H0:=-L+}
H_0 =  -   {\cal L} +\dot{x}^a P_a +  \dot{{\theta}}{}^\alpha \pi_\alpha + \dot{\rho}^\# P_{_{\rho^{\#}}} &+& f_\tau ^\# {\mathbf d}^{=}- f_\tau^= {\mathbf d}^{\#} - \omega_\tau {\mathbf d}^{(0)}- \qquad \nonumber \\ &-&  tr (\dot{{\bb X}} {\bb P}) + {i\over 4}  tr ( {\Psi}\dot{{\Psi}})  - tr (\dot{{\bb A}}_\tau {\bb P}_{_{{\bb A}}})  \; , \qquad
\end{eqnarray}
where $P_{_{\rho^{\#}}}=: P_{({\rho})\# } = \partial {\cal L}/ \partial \dot{\rho}^{\#}$ and ${\bb P}_{_{{\bb A}}}= \partial {\cal L}/ \dot{\partial {\bb A}}_\tau $ are the canonical momenta for the Lagrange multipliers: scalar density ${\rho}^{\#}$ and  1d  gauge field, which is a traceless $N\times N$ matrix ${\bb A}_\tau= \frac {\bb A} {d\tau}$ entering (\ref{DbbX=}).

Actually, as we will see in the next section, for our system the canonical Hamiltonian is equal to zero,
\begin{eqnarray}
\label{H0=0}
 H_0 = 0
\; , \qquad
\end{eqnarray}
which implies that the complete hamiltonian is a linear combination of (the primary and secondary first class) constraints. This can be expressed as
\begin{eqnarray}
\label{Hw=0}
H \approx 0
 \; , \qquad
\end{eqnarray}
 where $\approx $ denotes the {\it  weak equality}  \cite{Dirac:1963} i.e. equality which can be used only after all the Poisson brackets are calculated.

\subsection{Constraints  }

The calculation of the canonical/covariant momenta for the dynamical variables of the nAmW system results in the {\it primary constraints} (relations which involve coordinates and momenta but do not contain velocities  \cite{Dirac:1963})
\begin{eqnarray}
\label{tPhia=Pa-=0}
\tilde{\Phi}_a &:= & P_a -  {\rho}^\# u_a^= -  u_a^\#\, {1\over {\rho}^\# }\; {\cal H}  \approx 0 \; , \qquad \\
\label{dal=0}
d_\alpha  &:= & \pi_\alpha + i P_a (\gamma^a\theta)_\alpha - i \mu v_\alpha^+ {1\over \sqrt{{\rho}^\# }}  \; {\cal S}   \; \approx 0 \; , \qquad
 \\ \label{d--=0}
 && {\mathbf d}^{=}\approx 0 \; , \qquad
 \\ \label{d++=0}
 && {\mathbf d}^{\#}\approx 0 \; , \qquad
 \\ \label{td0=d0-=0}
 \tilde{{\mathbf d}}{}^{(0)}&:=& {\mathbf d}^{(0)} -2 {\rho^{\#}} P_{_{\rho^{\#}}} \approx 0 \; , \qquad
  \\ \label{Prho=J}
 && {\rho^{\#}} P_{_{\rho^{\#}}}-{\cal J} \approx 0 \; , \qquad
\\
\label{PAt=0} && {\bb P}_{_{{\bb A}}}\approx 0\; ,  \qquad
\end{eqnarray}
where  $ {\cal H}  $, ${\cal S}$ and ${\cal J}$ are defined in (\ref{HSYM=3D}) and (\ref{cJ=}).

The latter constraint highlights the Lagrange multiplier nature of the 1d SU(N) gauge field $ {\bb A}_\tau $. Indeed, the calculation of the canonical Hamiltonian with the use of the primary constraints gives $$H_0= tr\left( {\bb A}_\tau \left([{\bb P}, {\bb X}]- {i\over 4}\{ \Psi , \Psi  \}\right)\right). $$ The requirement of preservation of  the constraint (\ref{PAt=0}) in the evolution, $\dot{{\bb P}}_{_{{\bb A}}}= [{\bb P}_{_{{\bb A}}}, H_0] \approx 0$, results in the {\it secondary constraint}
\begin{eqnarray}
\label{bbG:=}
{\bb G}:=  [{\bb P}, {\bb X}]- {i\over 4}\{ \Psi , \Psi  \} \approx 0   \;  \qquad
\end{eqnarray}
which has the meaning of the Gauss law for the 1d gauge field.
Taking this into account, {as previously mentioned}, we find that the canonical Hamiltonian vanishes, $H \approx 0$ (see Eqs. (\ref{H0=0}) and (\ref{Hw=0})).

Notice that the Gauss law (\ref{bbG:=}) is the only essential  matrix constraint in our system (as (\ref{PAt=0}) is just a manifestation of the Lagrange multiplier nature of the 1d gauge field and other constraints do not carry uncontracted SU(N) indices).
Thus to write its brackets  with matrix variables, it is convenient to consider its trace with some probe matrix, which we denote by the same symbols ${\bb Y}$, ${\bb Z}$ as in (\ref{YPXZ=PB}) and (\ref{YPsiZPsi=}),
\begin{eqnarray}
\label{bbGX=PB}
[tr {\bb Y}{\bb G}, {\bb X}]_{P.B.} = [ {\bb X},  {\bb Y}]\; , \qquad  [tr {\bb Y}{\bb G}, {\bb P}]_{P.B.} = [ {\bb P},  {\bb Y}]\; , \qquad  [tr {\bb Y}{\bb G}, {\Psi}]_{P.B.} =  [ {\Psi}, {\bb Y}] \; . \qquad
\end{eqnarray}
These relations reflect the nature of the Gauss constraint as generator of  $SU(N)$ gauge symmetry. One can also check that this generator indeed obeys the $su(N)$ algebra on the Poisson brackets,
\begin{eqnarray}
\label{GG=PB=G}
[tr ({\bb Y}{\bb G}), tr ( {\bb Z}{\bb G})]_{P.B.} = tr ([{\bb Y},  {\bb Z}] {\bb G}) \; . \qquad
\end{eqnarray}
The Gauss constraint actually 'commutes' (this is to say has vanishing brackets) with all the other essential constraints of our system as all these are singlets of the $SU(N)$ gauge symmetry.

Taking into account the secondary constraint  (\ref{bbG:=}) we see that the current (\ref{HSYM=3D}) simplifies
\begin{eqnarray}
\label{cH=PP+XG}
&& {\cal H}= {\mu^2 \over 2} tr ( {\bb P} {\bb P} ) - 4i \mu^2 tr ( {\bb X} {\bb G} )\approx {\mu^2 \over 2} tr ( {\bb P} {\bb P} ):=  {\cal H}_0
   \; . \qquad
\end{eqnarray}
This, in its turn, allows to simplify the vector constraint (\ref{tPhia=Pa-=0})
\begin{eqnarray}
\label{Phia=Pa-=0}
\Phi_a &:= & P_a -  {\rho}^\# u_a^= -  u_a^\#\, {\mu^2\over 2{\rho}^\# }\;  tr ( {\bb P} {\bb P} )  \equiv P_a -  {\rho}^\# u_a^= -  u_a^\#\, {1\over {\rho}^\# }\;  {\cal H}_0  \approx 0 \;
   \; . \qquad
\end{eqnarray}
{For our purposes}, it will be convenient to split this latter into
\begin{eqnarray}
\label{Phi++=}
&& \Phi^\# := \Phi_a u^{a\#}=  P_a u^{a\#} - 2 {\rho}^\# \approx  0 \; , \qquad \\
\label{Phi--=}
&& \Phi^=:= \Phi_a u^{a=}:= P_a u^{a=} -  {\mu^2\over {\rho}^\# }\;  tr ( {\bb P} {\bb P} )  \equiv P_a u^{a=} -  {2\over {\rho}^\# }\;  {\cal H}_0  \approx 0 \;
   \;, \qquad \\
\label{PhiI=}
&& \Phi^\perp := \Phi_a u^{a\perp}\approx  0 \; . \qquad\end{eqnarray}

\subsection{Current algebra}

As our constraints involve the currents constructed from the matrix fields,  (\ref{cH=PP+XG}) and (\ref{cJ=}), it is useful to know their algebra and their Poisson  brackets with matrix fields. These are
\begin{eqnarray}
\label{cScS=PP}
&& \{ {\cal S},  {\cal S}\}_{_{PB}}= -2i tr ( {\bb P} {\bb P} )= -{4i\over \mu^2} {\cal H}_0  \; , \qquad \\
\label{cHcS=PsiG}
&& {} [{\cal H},  {\cal S}]_{_{PB}}=  0  \; , \qquad \\
\label{cJcS=-cS}
&& {} [{\cal J},  {\cal S}]_{_{PB}}= - {\cal S}   \; , \qquad \\ \label{cHcJ=XG}
&& {} [{\cal H}, {\cal J}]_{_{PB}}= \mu^2 tr ( {\bb P} {\bb P} ) =  2{\cal H}_0
   \; ,
\end{eqnarray}
and
\begin{eqnarray}
\label{cSX=}
&& {}[ {\cal S},  {\bb X}]_{_{PB}}=  {\Psi}  \; , \qquad \{ {\cal S},  {\Psi}\}_{_{PB}}= -2i {\bb P} \; , \qquad \\ \label{cJX=}
&& {}[ {\cal J},  {\bb X}]_{_{PB}}=  {\bb X}  \; , \qquad [ {\cal J},  {\bb P}]_{_{PB}}=  - {\bb P}  \; , \qquad [ {\cal J},  {\Psi}]_{_{PB}}= 0 \; . \qquad
\end{eqnarray}
The Gauss constraints obey
\begin{eqnarray}
\label{GcJ=0}
&& {}[ {\bb G}, {\cal J}  ]_{_{PB}}=  0   \; , \qquad {}[ {\bb G} , {\cal S}  ]_{_{PB}}=  0,   \;   \qquad
\end{eqnarray}
and has all the brackets with other constraints vanishing. This is due to the fact that all the essential constraints  are singlets under $SU(N)$ generated by the Gauss constraint.

Finally, let us also present the Poisson brackets of the fermionic constraints
(\ref{dal=0}),
\begin{eqnarray}
\label{dfdf=}
&& \{ d_\alpha ,  d_\beta \}_{_{PB}}= -4i \rho^\# v_\alpha^-v_\beta^- - 2i \Phi_a \gamma^a_{\alpha\beta}\approx -4i \rho^\# v_\alpha^-v_\beta^- \; . \qquad
\end{eqnarray}
Their form suggests that, as in the case of massless superparticle, $d_\alpha$ contains one first class constraint, $d^-:= d_\alpha v^{\alpha -}\approx 0$ and one second class constraint $d^+:= d_\alpha v^{\alpha +}\approx 0$.
{In the next section we will split   the first and second class  constraints of our dynamical system.}

\section{First class constraints, second class constraints and Dirac brackets}

Following Dirac \cite{Dirac:1963}, let us write the most generic ansatz fo the total Hamiltonian as a sum of constraints with Lagrange multipliers
\begin{eqnarray}
\label{HT=}
H=a^a\Phi_a+ \kappa^\alpha d_\alpha +\Lambda^\# {\bf d}^= - \Lambda^= {\bf d}^\# - \Lambda^{(0)}({\bf d}^{(0)}
-2 {\rho^{\#}} P_{_{\rho^{\#}}}) + \beta ({\rho^{\#}} P_{_{\rho^{\#}}}- {\frak{J}}) +tr({\bb Y}{\bb G}) \;  \qquad
\end{eqnarray}
and use it to check the preservation of the constraints
(\ref{Phia=Pa-=0}), (\ref{dal=0}), (\ref{d--=0}), (\ref{d++=0}), (\ref{td0=d0-=0}), (\ref{Prho=J}), (\ref{bbG:=}) (let us denote them ${\frak C}_{\frak{A}}\approx 0$)
under evolution ($\dot{{\frak C}}_{\frak{A}}= [{\frak C}_{\frak{A}}, H]_{P.B.}\approx 0$).

This procedure gives the expressions for a number of Lagrange multipliers,
\begin{eqnarray}
\label{LMs=}
\beta =0\; , \qquad  \Lambda^==-\Lambda^\# \frac{1}{ (\rho^\#)^2} {\cal H}_0\; ,\qquad \kappa^-:= \kappa^\alpha v_\alpha^-= \frac {\mu} {4(\rho^\#)^{3/2}}\Lambda^\# \cal{S}\; , \qquad
\\ \label{LMa=}
a^\perp := a^a u_a^\perp =\Lambda^\#{\cal P}_{\rho^\#} \; , \qquad a^= := a^a u_a^= = \frac{3a^\#}{2( \rho^\#)^2} {\cal H}_0 +\frac {3i\mu} {2(\rho^\#)^{3/2}}{\cal S}\kappa^+ \; , \qquad
\end{eqnarray}
which correspond to second class constraints.

Substituting the expressions (\ref{LMs=})  and (\ref{LMa=}) into  (\ref{HT=}), we find
\begin{eqnarray}
\label{HT==}
H=\frac 1 2 a^\# \left(\Phi^=  + \frac{3{\cal H}_0}{4( \rho^\#)^2}  \Phi^\#  \right) +  \kappa^+  \left(d^-  - \frac {3i\mu{\cal S}} {4(\rho^\#)^{3/2}}\Phi^\#   \right) - \qquad \nonumber \\ - \Lambda^{(0)} \left({\bf d}^{(0)}
-2 {\rho^{\#}} P_{_{\rho^{\#}}}\right) +  tr({\bb Y}{\bb G})+ \qquad  \nonumber \\
+ \Lambda^\# \left({\bf d}^= -
{\cal P}_{\rho^\#}\Phi^\perp -  \frac {\mu} {4(\rho^\#)^{3/2}}{\cal S}d^++ \frac 1 {( \rho^\#)^2} {\cal H}_0{\bf d}^\#\right) \; .
\end{eqnarray}
The Lagrange multipliers in the first two lines, bosonic $ a^\#$,   $\Lambda^{(0)}$ and ${\bb Y}$ and fermionic  $\kappa^+$, are arbitrary so that the expressions they multiply are the first class constraints generating the gauge symmetries of the system (see \cite{Dirac:1963}).

The remaining constraints are of the  second class:
\begin{eqnarray}
\label{IIclass=2}
{\bf d}^\#\approx 0 \; , \qquad \Phi^\perp= P^au_a^\perp  \approx 0 \; , \qquad \\
\label{IIclass=+2} \Phi^\#= P^au_a^\#- 2\rho^\#  \approx 0 \; , \qquad  {\rho^{\#}} P_{_{\rho^{\#}}} - {\cal J} \approx 0 \; , \qquad  \\
\label{IIclass=1f} d^+= v^{\alpha+}d_\alpha \approx 0 \; . \qquad
\end{eqnarray}
They have the following non vanishing Poisson brackets:
\begin{eqnarray}
\label{II-II=PB}
{} [{\bf d}^\#, \Phi^\perp]_{P.B.}=  \Phi^\#+ 2\rho^\#  \approx  2\rho^\#  \; , \qquad [{\rho^{\#}} P_{_{\rho^{\#}}} - {\cal J}\, , \;
\Phi^\# ]_{P.B.}=  2\rho^\#  \; , \qquad
\end{eqnarray}
and \begin{eqnarray}
\label{dfdf=}
&& \{ d^+ ,  d^+ \}_{_{PB}}= -4i \rho^\#  - 2i \Phi^{\#} \approx -4i \rho^\#  \; . \qquad
\end{eqnarray}
This allows to define the Dirac brackets
\begin{eqnarray}
\label{DB=PB+}
&{}  [ ... , ...\}_{_{DB}}&= \qquad \nonumber \\ & = [ ... , ...\}_{_{PB}}& - [ ... ,  \Phi^\perp\}_{_{PB}} \frac {1} {\Phi^\#+ 2\rho^\#} [{\bf d}^\# , ...\}_{_{PB}}  +  [ ... , {\bf d}^\#\}_{_{PB}} \frac {1} {\Phi^\#+ 2\rho^\#} [  \Phi^\perp , ...\}_{_{PB}} - \nonumber \\  &&- [ ... , \Phi^\# \}_{_{PB}} \frac {1} {2\rho^\# } [ {\rho^{\#}} P_{_{\rho^{\#}}} - {\cal J}\,  , ...\}_{_{PB}}+  [ ... , {\rho^{\#}} P_{_{\rho^{\#}}} - {\cal J}\, \}_{_{PB}} \frac {1} {2\rho^\# } [ \Phi^\#  , ...\}_{_{PB}}  -\qquad \nonumber \\ &&  - [ ... , d^+\}_{_{PB}} \frac {i} {2(\Phi^\#+ 2\rho^\#)} [ d^+ , ...\}_{_{PB}}\; .
\end{eqnarray}
If using these {Dirac brackets}, we can treat the second class constraints  (\ref{IIclass=2}), (\ref{IIclass=+2}), (\ref{IIclass=1f}) as strong equalities. However, we will use a different method of dealing with the second class constraints.

\section{Generalized Gupta-Bleuler method/conversion of the second class constraints}

Instead of using the Dirac bracket we prefer to apply the generalized Gupta-Bleuler method to the bosonic second class constraints and the conversion method \cite{Batalin:1986fm,Batalin:1989dm,Egorian:1993sc}  for the fermionic second class constraints to arrive at the classically equivalent system with the first class constraints only.
The generalized Gupta--Bleuler method (see \cite{Bandos:1999qf}), which is actually equivalent to conversion method of \cite{Batalin:1986fm,Batalin:1989dm,Egorian:1993sc},  consists in omitting one of two conjugate second class constraints. Then the remaining constraints of the canonically conjugate pair becomes the first class and generate some gauge symmetry of the modified dynamical system\footnote{In the original Gupta-Bleuler method (see e.g. \cite{de Azcarraga:1982dw} and refs. therein) the the pair of second class constraints, which one subject to such a procedure,  should be conjugate not only with respect to Poisson brackets, but also related by complex or hermitian conjugation. The generalized Gupte-Bleuler procedure can be obtained by conversion, see e.g. \cite{Bandos:1999qf}.}.

In our case we prefer to omit the first constraint in the pair (\ref{IIclass=2}) and the second in (\ref{IIclass=+2}). Then
the constraints $\Phi^\perp\approx 0$ and $\Phi^\#\approx 0$ become first class and generate some gauge symmetries. The original conjugate second class constraints ${\bf d}^\#$ and ${\rho^{\#}} P_{_{\rho^{\#}}} - {\cal J} $ can be treated as gauge fixed conditions for such gauge symmetry. Fixing these condition we arrive at the original description of the dynamical system.

As far as the fermionic second class constraint is concerned, we will convert it to  first class by introducing the special conversion degree of freedom, the fermionic variable $\xi$ which is its own momentum. This is to say it satisfies the Poisson bracket relation
\begin{eqnarray}
\label{xixi=PB}
&& \{ \xi ,  \xi \}_{_{PB}}= -i  \; . \qquad
\end{eqnarray}
We use this variable to modify the fermionic constraint (\ref{dal=0}) to
\begin{eqnarray}
\label{tdal=I}
\tilde{d}_\alpha  &:= & d_\alpha -2 v_\alpha^- i\sqrt{{\rho}^\# }\, \xi \qquad \nonumber
\\ &=& \pi_\alpha + i P_a (\gamma^a\theta)_\alpha - i \mu  v_\alpha^+ \; {\cal S}/\sqrt{{\rho}^\# } -2  i\sqrt{{\rho}^\# }\, v_\alpha^- \, \xi \approx 0 \; .
 \qquad
\end{eqnarray}
We also use this new constraint to modify one of the bosonic first class constraints (by setting in it $d^+\approx -2 \sqrt{\rho^\#}\xi$) so that the  complete set of the essential bosonic constraints includes
\begin{eqnarray}
\label{Phia=I}
{\Phi}_a &:= & P_a -  {\rho}^\# u_a^= -  u_a^\#\, {1\over {\rho}^\# }\; {\cal H}_0  \approx 0 \; , \qquad \\
\label{td--=I}
\tilde{{\bf d}}{}^= &:=& {\bf d}^= -  \frac 1 {( \rho^\#)^2} {\cal H}_0{\bf d}^\#
 - i \frac {\mu} {2\rho^\#}{\cal S}\xi \approx 0 \; , \qquad \\
\label{td0=I}\tilde{{\bf d}}^{(0)} &:=& {\bf d}^{(0)}
-2 {\rho^{\#}} P_{_{\rho^{\#}}} \approx 0 \; , \qquad \\ \label{bbG=I}
{\bb G} &:=&  [{\bb P}, {\bb X}]- {i\over 4}\{ \Psi , \Psi  \} \approx 0.
\end{eqnarray}
The only nonvanishing brackets in the algebra of the first class constraints (\ref{tdal=I})--(\ref{bbG=I}) are
\begin{eqnarray}
\label{tdtd=Phi}
{} \{ \tilde{d}_\alpha, \tilde{d}_\beta\}_{_{P.B.}}&=& -2i \Phi_a \gamma^a_{\alpha\beta}\; , \qquad
\\
\label{td0td--=-2td--}
{} [\tilde{{\bf d}}^{(0)}, \tilde{{\bf d}}{}^=]_{_{P.B.}}  &=&-2 \tilde{{\bf d}}{}^=\; .
\end{eqnarray}
Thus, these constraints are of the first class and their quantum counterparts can be imposed as conditions on the state vector of the quantized system.

\section{Quantization of the converted dynamical system and the  equations of a  superfield theory of 3D Matrix model}

The quantization implies substituting canonical variables of the dynamical system by operators acting on the state vectors. The commutators/anticommutators  of the operators can be reproduced from the Poisson brackets of the classical variables by the famous Dirac prescription
\begin{eqnarray}
\label{PB->com}
&& [ ... , ... \}_{_{PB}}\; \mapsto \; \frac 1 {i\hbar }  [ ... , ... \} \; . \qquad
\end{eqnarray}
In particular this rule gives for quantum version of our fermionic conversion variable $\xi$
\begin{eqnarray}
\label{hxihxi=1}
&& \{ \hat{\xi} ,  \hat{\xi} \}= 1  \; . \qquad
\end{eqnarray}

\if{}
Hence it can be realized by considering the state vector to be a doublet and to use the matrix realization
\begin{eqnarray}
\label{hxi=mat}\hat{\xi}=\frac 1 {\sqrt{2}}\left(\begin{matrix} 0 & 1 \cr   1 & 0 \end{matrix} \right)\qquad
\end{eqnarray} ({\it cf.} \cite{Berezin:1976eg,Gershun:1979fb}).
\fi

Similarly, the quantum version of the internal fermionic variables obeys
 \begin{eqnarray}
\label{PsiPsi=}
 {}\{ \hat{\Psi}_i{}^j,  \hat{\Psi}{}_{i'}{}^{j'}\} = 2\left(\delta _i{}^{j'}\delta_{i'}{}^j -
 {1\over N}
 \delta_i{}^j\delta_{i'}{}^{j'} \right) \; , \qquad i,j,i', j'=1,...,N\; .
\qquad
\end{eqnarray}
\if{}
Notice an interesting interplay between matrix indices and position of the operators which result in
 \begin{eqnarray}
\label{trPsiPsi=}
 {}\{\hat{\Psi}_i{}^k,  \hat{\Psi}{}_{k}{}^{j}\} = 2\left(N-
 {1\over N}\right) \delta _i{}^{j} \; , \qquad {}\hat{\Psi}_i{}^j \hat{\Psi}{}_{j}{}^{i}= \frac 1 2 \{\hat{\Psi}_i{}^j,  \hat{\Psi}{}_{j}{}^{i}\} = (N^2-1)\; .
\qquad
\end{eqnarray}
\fi

For $N=2$, the fermionic operators can be decomposed on Pauli matrices,
 \begin{eqnarray}
\label{Psi=PsiIsI}
 \hat{\Psi}_i{}^j= \frac 1 {\sqrt{2}}  \hat{\Psi}{}^I\sigma^I_{i}{}^{j}\; , \qquad
\end{eqnarray}
and (\ref{PsiPsi=}) implies that operators $\hat{\Psi}{}^I$ obey  3-dimensional Clifford algebra
\begin{eqnarray}
\label{PsiIPsiJ=}
{}\{   \hat{\Psi}{}^I,  \hat{\Psi}{}^J \} = 2 \delta^{IJ} \; . \qquad
\end{eqnarray}

Actually, the algebra (\ref{PsiPsi=}) is isomorphic to $(N^2-1)$ dimensional Clifford algebra for an arbitrary $N$. This follows from the completeness relations for the SU(N) generators 
\begin{eqnarray}
\label{TATA=}
T^A{}_i{}^j T^A{}_{i'}{}^ {j'}=  \delta _i{}^{j'}\delta_{i'}{}^j -
 {1\over N}
 \delta_i{}^j\delta_{i'}{}^{j'} \; , \qquad A=1,...,N^2-1\; , \qquad i,j,i',j'=1,...,N\; . \qquad
\end{eqnarray}
The traceless $N\times N$ matrix operators can be expressed by 
 \begin{eqnarray}
\label{Psi=PsiATA}
 \hat{\Psi}_i{}^j=  \hat{\Psi}{}^A T^A{}_{i}{}^{j}\; , \qquad
\end{eqnarray}
in terms of fermionic operators  $\hat{\Psi}{}^A$ obeying the Clifford algebra
\begin{eqnarray}
\label{PsiAPsiB=}
{}\{   \hat{\Psi}{}^A,  \hat{\Psi}{}^B \} = 2 \delta^{AB} \; , \qquad A,B=1,...,N^2-1\; . \qquad
\end{eqnarray}
However, we have preferred to introduce a separate notation for the simplest nontrivial case of N=2 as just for this case   
 we will discuss below in detail an explicit realization of the Clifford generators on the state vectors.

The rest of the operators can be taken in coordinate or momentum representation, for instance
 \begin{eqnarray}
\label{coord-repP}
\hat{P}_a&=&-i\partial_a, \qquad \hat{\pi}_\alpha = - i \partial_\alpha, \qquad \hat{d}^== iD^=\; , \quad  \hat{d}^\#= iD^\#\; ,
\qquad \nonumber \\
{\cal P}_{\rho^\#}&=&
-i \frac \partial {\partial \rho^\#},\qquad \hat{{\bb X}}_i{}^j = -i \frac \partial {\partial {\bb P}_j{}^i }\; , \qquad \\ \label{coord-repX}
\hat{x}^a&=&x^a\; , \qquad \hat{\theta}^\alpha={\theta}^\alpha\; , \qquad \hat{v}_\alpha^\mp ={v}_\alpha^\mp \; , \qquad
\hat{\rho}^\# = {\rho}^\# \; , \qquad
 \hat{{\bb P}}_i{}^j = {\bb P}_i{}^j\; .
\qquad
\end{eqnarray}
In (\ref{coord-repP})  and below
\begin{eqnarray}
\label{D=:=} D^== v_\alpha^-\frac \partial {\partial v_\alpha^+ }\; , \qquad D^\#= v_\alpha^+\frac  \partial {\partial v_\alpha^- }\; , \qquad
D^{(0)}=  v_\alpha^+\frac \partial {\partial v_\alpha^+ }-  v_\alpha^-\frac  \partial {\partial v_\alpha^-}
\qquad
\end{eqnarray}
are covariant derivatives of Lorentz harmonic variables (see (\ref{bbD++})) and
\begin{eqnarray}
\label{Dal:=}
D_\alpha:= \partial_\alpha +i (\gamma^a\theta)_\alpha \partial_a
\qquad
\end{eqnarray}
 is the fermionic covariant derivative which obeys
\begin{eqnarray}
\label{DD=}
{} \{D_\alpha,D_\beta\} = 2i \gamma^a_{\alpha\beta} \partial_a\; .
\qquad
\end{eqnarray}

\bigskip

For the generic value of $N$, the quantum first class constraints imposed on the state vector $\Xi$, which represented by a function of variables (\ref{coord-repX})  are
\begin{eqnarray}
\label{partXi=}
&& \left( \partial_a -  i{\rho}^\# u_a^= -  {i\over {\rho}^\# }\; u_a^\#\, {\cal H}_0 \right) \Xi = 0 \; , \qquad \\
\label{D--Xi=} &&
\left( D^= -  \frac 1 {( \rho^\#)^2} {\cal H}_0 D^\#
 -  \frac {\mu} {2\rho^\#}\hat{{\cal S}}\hat{\xi}\right) \Xi = 0 \; , \qquad \\
\label{D0Xi=}
&& \left( D^{(0)}
+2 {\rho^{\#}} \frac \partial {\partial \rho^\#} \right) \Xi = 0 \; , \qquad \\
\label{bbGXi=}
i\hat{{\bb G}}_i{}^j \Xi &:=& \left({\bb P}_i{}^k  \frac \partial {\partial {\bb P}_j{}^k } -  \frac \partial {\partial {\bb P}_k{}^i } {\bb P}_k{}^j + {1\over 2}\hat{\Psi}_i{}^k \hat{\Psi}_k{}^j + \frac {(N^2-1)} {2N}\delta_i{}^j  \right) \Xi = \; , \qquad \nonumber \\
&=& \left({\bb P}_i{}^k  \frac \partial {\partial {\bb P}_j{}^k } -  {\bb P}_k{}^j  \frac \partial {\partial {\bb P}_k{}^i } + {1\over 4}[\hat{\Psi}_i{}^k, \hat{\Psi}_k{}^j]  \right) \Xi = 0 \; , \qquad
\\ \label{DXi+=}
i\hat{\tilde{d}}_\alpha \Xi &:=& \left(D_\alpha + \frac {\mu} {\sqrt{\rho^\#}} v_\alpha^+ \hat{{\cal S}} +
2\sqrt{\rho^\#} v_\alpha^- \hat{\xi}\right) \Xi = 0\; . \qquad \end{eqnarray}
Here ${\cal H}_0={\mu^2 \over 2} tr ( {\bb P} {\bb P} )$ (see (\ref{cH=PP+XG})) and (see (\ref{cJ=}))
\begin{eqnarray}
\label{hatS:=} \hat{{\cal S}}=tr (\hat{\Psi}{\bb P}) \; . \qquad
\end{eqnarray}

\if{}
As we have discussed above,
to use the matrix representation of Clifford algebra valued fermionic variables we have to allow the state vector carries  indices, see   (\ref{Xi=8}), (\ref{cBA=}), (\ref{PsiIfFA=}) for $N=2$ case. We have written the system of equations in an abstract form which allow a possibility of different representations for the fermionic operator.
\fi

Notice the presence of an ordering constant in the first expression for quantum $su(N)$ generator (\ref{bbGXi=}). Its value is fixed by the requirement that the operator $\hat{{\bb G}}_i{}^j$ is traceless on its matrix indices, $\hat{{\bb G}}_i{}^i=0$.
This constant contains two contributions one of which comes from the fermionic operators (and with opposite sign), so that in the bosonic limit this constant should be multiplied by 2.  Actually the constants can be set to zero by putting derivative on the right and writing ${1\over 4}\hat{\Psi}_i{}^k \hat{\Psi}_k{}^j- {1\over 4} \hat{\Psi}_k{}^j\hat{\Psi}_i{}^k $
instead of ${1\over 2}\hat{\Psi}_i{}^k \hat{\Psi}_k{}^j$. This is indicated in the second form of Eq. (\ref{bbGXi=}).


It is convenient to decompose Eqs. (\ref{partXi=}) on three components
\begin{eqnarray}
\label{d++Xi=}
 \partial^\# \Xi = 2 i{\rho}^\# \Xi\; , \qquad  \partial^\# := u^{\# a} \partial_a  \; , \qquad  \\
 \label{d--Xi=}
 \partial^= \Xi =  {2i\over {\rho}^\# }\;  {\cal H}_0  \Xi\; , \qquad  \partial^= := u^{= a} \partial_a  \; , \qquad \\
 \label{dperpXi=}
 \partial^\perp \Xi = 0\; , \qquad  \partial^\# := u^{\perp a} \partial_a  \; . \qquad
\end{eqnarray}
Then, splitting the fermionic covariant derivatives (\ref{Dal:=})
\begin{eqnarray}
\label{D-:=}
D^+= v^{\alpha +}D_\alpha\; , \qquad D^-= v^{\alpha -}D_\alpha\; , \qquad
\end{eqnarray}
we find that their superalgebra (\ref{DD=}), when calculated on the state vector with taking into account (\ref{d++Xi=})--(\ref{dperpXi=}), simplifies to
\begin{eqnarray}
\label{D-D-=}
D^+D^+\Xi= -2{\rho}^\# \Xi \; , \qquad  D^-D^-\Xi = -{2\over {\rho}^\# }\;  {\cal H}_0  \Xi\; , \qquad
\{ D^+,  D^-\} \Xi=0\; . \qquad
\end{eqnarray}
Furthermore, one can check that   (\ref{partXi=}) or (\ref{d++Xi=})--(\ref{dperpXi=}) imply  that the action of split derivatives on the state vector can be described by
\begin{eqnarray}
\label{D+Xi=d+Xi+}
D^+\Xi &=& (\partial^+ + 2\theta^- {\rho}^\#)  \Xi =  (-\partial_-+ 2\theta^- {\rho}^\#)  \Xi  \; , \qquad  \\
\label{D-Xi=d-Xi+}
D^-\Xi &=& \left(\partial^- - \theta^+ {2\over {\rho}^\# }\;  {\cal H}_0 \right)  \Xi = \left(\partial_+ - \theta^+ {2\over {\rho}^\# }\;  {\cal H}_0 \right)  \Xi   \; , \qquad
\end{eqnarray}
where
\begin{eqnarray}
\label{th+=thv+}
\theta^\mp =\theta^\alpha v_\alpha^\mp\; , \qquad \partial^\pm = v^{\alpha \pm}\partial_\alpha\; , \qquad \\
\label{part-=ddth-}
\partial_\mp = \frac {\partial} {\partial \theta^\mp}= \mp \partial^\pm \; . \qquad
\end{eqnarray}

Using the fermionic derivatives (\ref{D-:=}), we can write Eq.  (\ref{DXi+=}) as system of two equations
\begin{eqnarray}
\label{xiXi=}
\hat{\xi}\Xi = \frac 1 {2\sqrt{{\rho}^\# }}D^+\Xi\; , \qquad \\
\label{D-Xi=}
\left(\hat{{\cal S}}+ \frac {\sqrt{{\rho}^\# }}{\mu} D^- \right)\Xi=0\; . \qquad
\end{eqnarray}
The first of these equations, (\ref{xiXi=}) completely determines the action of the quantum Clifford variable $\hat{\xi}$ on the state vector. We can use this equation to modify (\ref{D--Xi=}) and then do not consider more this  $\hat{\xi}$ operator, at least when analysing the quantum constraints.

Furthermore, using (\ref{D-Xi=}) we can exclude also the fermionic matrix operator $\Psi_i{}^j$ from (\ref{D--Xi=}) which then reads
\begin{eqnarray}
\label{D--Xi=D+D-} &&
\left( D^= -  \frac 1 {( \rho^\#)^2} {\cal H}_0 D^\#
 -  \frac {1} {4\rho^\#}D^+D^-\right) \Xi = 0 \; . \qquad
\end{eqnarray}
Then the  fermionic matrix operator $\Psi_i{}^j$ remains involved, besides (\ref{D-Xi=}),  only in the Gauss low constraint (\ref{bbGXi=}) which implies that the state vector is singlet under the $SU(N)$ symmetry.

\bigskip

To argue in favour of that one can use (\ref{xiXi=}) to completely exclude the Clifford variable $\hat{\xi}$ from the consideration, we can consider realization of one dimensional Clifford algebra on Clifford superfields (see
\cite{Sorokin:1989jj,Bandos:1999qf}). Namely, let us consider the state vector to be a function of the Clifford algebra valued variable   $\hat{\xi}$ itself. Then, as $\hat{\xi}^2=1/2$, the decomposition of the state vector on this fermionic variables will contain two terms only,
$$
\Xi=\Sigma_0 +i\hat{\xi} \Sigma_1\; ,
$$
and equation (\ref{xiXi=}) just provides us with the expression for $\Sigma_1$ in terms of $\Sigma_0$, \footnote{Besides (\ref{S1=DS0}),  (\ref{xiXi=})  produces $\Sigma_0= - \frac i {2\sqrt{\rho^\#}}D^+\Sigma_1$, but this is satisfied identically due to (\ref{S1=DS0}) and (\ref{D-D-=})
}
\begin{eqnarray}
\label{S1=DS0}
\Sigma_1= - \frac i {\sqrt{\rho^\#}}D^+\Sigma_0
\; . \qquad
\end{eqnarray}
Thus the quantum system is completely described in terms of $\hat{\xi}$-independent wavefunction
$\Sigma_0$  obeying  (\ref{partXi=}), (\ref{D0Xi=}), (\ref{bbGXi=}), (\ref{D--Xi=D+D-}) and (\ref{D-Xi=}).

Similar realization for the internal fermionic operators
$\Psi_i{}^j$ can be used in  $N=2$ case, where the decomposition (\ref{Psi=PsiIsI}) and $\Psi^I$ obey the
$d=3$ Clifford algebra (\ref{PsiIPsiJ=}) and, hence,
\begin{eqnarray}
\label{PsiXPsi=}
\Psi^I\Psi^J= \delta^{IJ}+ \epsilon^{IJK}\Psi^{\wedge 2}_K\; , \qquad \\
\label{PsiXPsi2=}
\Psi^I\Psi^{\wedge 2}_J= \delta^{IJ}\Psi^{\wedge 3} - \epsilon^{IJK}\Psi^K\; , \qquad  \\ \label{PsiXPsi3=}
\Psi^I\Psi^{\wedge 3}= \Psi^{\wedge 2}_I
\; ,  \qquad
\end{eqnarray}
where
\begin{eqnarray}
\label{Psi2I=}
\Psi^{\wedge 2}_I:=\frac 1 2 \epsilon^{IJK}\Psi^{J}\Psi^{K}
\; ,  \qquad \\
\label{Psi3=}
\Psi^{\wedge 3}:=\frac 1 {3!} \epsilon^{IJK}\Psi^{I}\Psi^{J}\Psi^{K}
\; .  \qquad
\end{eqnarray}

We can consider the state vector (component) $\Sigma_0$ to be a function of the 3 Clifford variables and decompose it in series which is clearly finite,
\begin{eqnarray}
\label{Sigma0=}\Sigma_0 (\Psi) = \phi^{(0)} + \Psi^I \phi^{(1)}_I + \Psi^{\wedge 2}_I \phi^{(2)}_I + \Psi^{\wedge 3} \phi^{(3)}
\; .  \qquad
\end{eqnarray}
It is not difficult to find that the Clifford algebra valued operator acts on this superfield as
\begin{eqnarray}
\label{PsiSigma0=}\Psi^I \Sigma_0  =  \phi^{(1)}_I +   \Psi^J (\phi^{(0)}\delta^{IJ}+ \epsilon^{IJK}\phi^{(2)}_K ) + \Psi^{\wedge 2}_J
(\phi^{(3)}\delta^{IJ}- \epsilon^{IJK}\phi^{(1)}_K )
\Psi^{\wedge 3}
\phi^{(2)}_I \; .   \qquad
\end{eqnarray}
Using this we can calculate the first term in (\ref{D-Xi=}) which in the case of $N=2$ includes the operator
\begin{eqnarray}
\label{hatS:=N2} N=2:\qquad  \hat{{\cal S}}=tr (\hat{\Psi}{\bb P})= \frac 1 {\sqrt{2}} \hat{\Psi}^I  tr(\sigma^{I}{\bb P})=
\hat{\Psi}^I p^I,
\end{eqnarray}
where
\begin{eqnarray}
\label{pI:=}  p^I:=  \frac 1 {\sqrt{2}} tr(\sigma^{I}{\bb P}).
\end{eqnarray}
In this way we split Eq.  (\ref{D-Xi=}) for $\Sigma_0$ into the following set of equations for its components defined in (\ref{Sigma0=}):
\begin{eqnarray}
\label{D-phi0=}  D^- \phi^{(0)} &=& - \frac \mu {\sqrt{\rho^\#}}p^I \phi^{(1)}_I\; , \qquad \\
\label{D-phi1=}  D^-  \phi^{(1)}_I&=&  \; \frac \mu {\sqrt{\rho^\#}} \left(p^I \phi^{(0)}-  \epsilon^{IJK}p^J \phi^{(2)}_K \right)\; , \qquad \\
\label{D-phi2=}  D^-  \phi^{(2)}_I&=& - \frac \mu {\sqrt{\rho^\#}} \left(p^I \phi^{(3)}+ \epsilon^{IJK}p^J \phi^{(1)}_K \right)\; , \qquad \\
\label{D-phi3=}  D^- \phi^{(3)} &=& \;  \frac \mu {\sqrt{\rho^\#}}p^I \phi^{(2)}_I\; . \qquad
\end{eqnarray}
When $p^I\not=0$, these system of equations can be solved by expressing, say  $\phi^{(1)}_I$ and $\phi^{(3)}$ in terms of $D^-$ derivatives of
 $\phi^{(2)}_I$ and $\phi^{(0)}$. When $p^I=0$ the equations imply just that the Clifford superfield $\Sigma_0$ is in kernel of the $D^-$ derivative,
 $D^-\Sigma_0=0$ and so are all its components.

A similar representation exists also for higher $N$. In it the state vector will be a Clifford superfield 
$\Sigma_0(\Psi^A)=\phi^{(0)}+\Psi^A\phi^{(1)}_A+ \ldots  + \Psi^{\wedge (N^2-1)}\phi^{(N^2-1)}$ with $N^2$ components carrying different representations of SU(N). Already for N=3 it will be decomposition in 1+9+36+84+126+126+84+36+9+1 representations of SU(N), which gives the idea on how non-minimal such a representation of the Clifford algebra  is. 

We do not study further this Clifford superfield description of the state vector of our nAmW system in this sense leaving  
our quantization on the level of ''proof of concept''.  
The reason is that the simplest in many respect $d=3$ case we are studying in this paper has a very specific problem of dealing with single Clifford algebra valued operator $\hat{\xi}$ and single fermionic matrix field
$\Psi_i{}^j$, and just this pushes us to the Clifford superfield formalism. All the higher dimensional multiwave systems, beginning $D=4$ nAmW (still to be constructed) and including $D=11$ mM0 system, will have an even number of both variables. Then both these sets can be split (in an $SU(N)$ covariant manner) on two conjugate subsets which can be considered as the sets of  coordinates and momenta (or of creation and annihilation operators). Hence  the system can be quantized in a coordinate (holomorphic) representation with respect to these operator variables. Thus the need to consider Clifford superfields or other exotic realizations of the operators will not appear in these more interesting higher dimensional cases.


To summarize, in this section we have performed the operator quantization of the converted version of our dynamical system, $d=3$ nAmW, and found the system of equation which are imposed on the quantum state vector of the system as a result of such quantization.
This system including (\ref{partXi=}), (\ref{D0Xi=}), (\ref{bbGXi=}), (\ref{D--Xi=D+D-}) and (\ref{xiXi=}), (\ref{D-Xi=}) can be called
equations for the field theory of the D=3 supersymmetric matrix model. One can say that this three dimensional supersymmetric field theory is defined on  a prototype of a `non-commutative space', namely on the space with matrix coordinates. In our simple case, the configuration space contains only  one bosonic matrix coordinate. 
We have written it in a form with matrix coordinate of the configuration  space being Fourier transformed, so that the non-commutative component is in momentum space. In {the} higher dimensional $D\geq 4$ counterparts of our  system, the sector with matrix coordinates will contain $2$ or more elements so that
there will be more reasons to speak about non-commutativity.


We have described the  explicit realization of the
fermionic variables $\hat{\xi}$ and $\hat{\Psi}_i{}^j$ on the state vector by considering this as a Clifford superfield. For the simplest case of 
$N=2$ we have elaborated this description in a bit more detail, while for higher $N$ it looks quite nonminimal.   Notice that it is tempting, following
\cite{Berezin:1976eg,Gershun:1979fb} to use the  matrix representations of the algebras (\ref{hxihxi=1}) and (\ref{PsiAPsiB=}) and to supply the state vector with indices corresponding to this matrix representations. 
However, the presence of Grassmann fermionic variables $\theta^\alpha$ and the realization of the constraints with the use of differentail operator in the space of these variables (this is to say in Grassmann algebra) makes the consistency of the use of matrix representation for the other, Clifford-type fermionic variables, not manifest.

We finish this study by
checking that the purely bosonic limit of the above system of equations for the field theory of D=3 non-Abelian multiwave makes sense, {namely by showing}  that the equations have a nontrivial solution in this limit.


\section{A solution of the bosonic limit of our Matrix field theory equations}

Setting all fermions equal to zero in a consistent way, we find the following set of equations imposed on the scalar wave function
$\Xi_0=\Xi_0(x^a, v_\alpha^\mp, {\rho}^\#, {\bb P})$:
\begin{eqnarray}
\label{partXi=b}
&& \left( \partial_a -  i{\rho}^\# u_a^= -  {i\over {\rho}^\# }\; u_a^\#\, {\cal H}_0 \right) \Xi_0 = 0 \; , \qquad \\
\label{D--Xi=b} &&
\left( D^= -  \frac 1 {( \rho^\#)^2} {\cal H}_0 D^\#\right) \Xi_0 = 0 \; , \qquad \\
\label{D0Xi=b}
&& \left( D^{(0)}
+2 {\rho^{\#}} \frac \partial {\partial \rho^\#} \right) \Xi_0 = 0 \; , \qquad \\ \label{bbGXi=b}
i\hat{{\bb G}}_i{}^j \Xi &:=& \left({\bb P}_i{}^k  \frac \partial {\partial {\bb P}_j{}^k } -  {\bb P}_k{}^j \frac \partial {\partial {\bb P}_k{}^i }\right) \Xi_0 = 0\; . \qquad
\end{eqnarray}
The constraint (\ref{partXi=b}) is solved by
\begin{eqnarray}
\label{Xi=expXi}
 \Xi_0 = e^{{i{\rho}^\# x^= + \frac i {{\rho}^\#} x^\# {\cal H}_0}}\Upsilon( v_\alpha^\mp, {\rho}^\#, {\bb P})
\; . \qquad
\end{eqnarray}
From now on we use
\begin{eqnarray}
\label{x--:=}
x^== x^au_a^= \, , \qquad x^\#= x^au_a^\# \, , \qquad x^\perp = x^au_a^\perp \, , \qquad
\end{eqnarray}
where the moving frame vectors are constructed from spinor moving frame variables according to  (\ref{u--=3D}), (\ref{u++=3D}), (\ref{up=3D}).

The constraint (\ref{bbGXi=b}) implies that $\Upsilon$ and $\Xi_0$ depend on the singlet combinations of the matrix momenta,
$ tr ({\bb P}{\bb P})= 2{\cal H}_0/\mu^2$,  $ tr ({\bb P}^3)$, $\ldots$, $ tr ({\bb P}^n)$, $\ldots$.
 Equation (\ref{D0Xi=b}) implies that $\Upsilon$ and $\Xi_0$ has zero weight with respect to $SO(1,1)$ symmetry acting on $v_\alpha^\mp$ and
 ${\rho}^\#$ in way indicated by their sign indices.

 To analyse (\ref{D--Xi=b}), it is important to notice that $$D^=  e^{{i{\rho}^\# x^= + \frac i {{\rho}^\#} x^\# {\cal H}_0}}= e^{{i{\rho}^\# x^= + \frac i {{\rho}^\#} x^\# {\cal H}_0}}\left(D^= + \frac {2ix^\perp} {{\rho}^\#}  {\cal H}_0 \right)\; , \qquad $$
 $$D^\#  e^{{i{\rho}^\# x^= + \frac i {{\rho}^\#} x^\# {\cal H}_0}}= e^{{i{\rho}^\# x^= + \frac i {{\rho}^\#} x^\# {\cal H}_0}}\left(D^\# + 2ix^\perp {{\rho}^\#} \right)\; , \qquad $$
 so that $\left( D^= -  \frac 1 {( \rho^\#)^2} {\cal H}_0 D^\#\right)$ commutes with $e^{{i{\rho}^\# x^= + \frac i {{\rho}^\#} x^\# {\cal H}_0}}$.
 Thus (\ref{D--Xi=b}) implies
 \begin{eqnarray}
\label{D--Y=b} &&
 D^= \Upsilon =   \frac 1 {( \rho^\#)^2} {\cal H}_0 D^\# \Upsilon  \; , \qquad
\end{eqnarray}
 where ${\cal H}_0={\mu^2 \over 2} tr ( {\bb P} {\bb P} )$ (\ref{cH=PP+XG}).

At ${\cal H}_0=0$, which implies ${\bb P}_i{}^j=0$, Eq. (\ref{D--Y=b}) reduces to
\begin{eqnarray}
\label{D--Y0=0} &&
 D^= \Upsilon_0 = 0\; .
\end{eqnarray}
The general solution of this equation is given by $\Upsilon_0$ independent on
$v_\alpha^+$, i.e. $\Upsilon_0= \Upsilon_0  (v_\alpha^-, {\rho}^\#)$ (see \cite{Delduc:1989ah} and \cite{Galperin:2001uw}). This is characteristic for the wave function describing the quantum states of a  massless (super)particle (see e.g. \cite{Bandos:1999qf}), which is a consistent result as, when
seting all our 'relative' momenta variables of multiparticle system equal to zero, our system reduces to the massless superparticle.

{Now we proceed to the} study of the solution for nonzero matrix momenta, in the branch where ${\cal H}_0\not=0$. To this end we decompose the wave function in power series on the spinor frame variables  \footnote{Such a decomposition implies passing to Euclidean version of the theory. The decomposition in the case of
Minkowski spacetime is more complicated, see e.g. \cite{B90,Fedoruk:1994ij} for 4d case. However, for our discussion aimed to show the presence of nontrivial solution,  it is sufficient, following \cite{Delduc:1989ah},  to study the solutions in the class of real analytic functions of spinor frame variables.
} $v_\alpha^\mp$,
\begin{eqnarray}
\label{Y=an} &&
  \Upsilon = \sum\limits_{n=0}^\infty \sum\limits_{m=0}^\infty  ( \rho^\#)^{\frac {n-m} 2}v_{(\alpha_1}^- \ldots v_{\alpha_n}^-v_{\beta_1}^+ \ldots v_{\beta_m)}^+\;  \phi_{n,m}^{\alpha_1\ldots\alpha_n\beta_1\ldots \beta_m}\; .
    \qquad
\end{eqnarray}
Substituting these into equation (\ref{D--Y=b}), we find the generic relation between coefficients
\begin{eqnarray}
\label{phi-1+1=phi+1-1} &&
(m+1)  \phi_{n-1,m+1}^{\alpha_1\ldots\alpha_n\beta_1\ldots \beta_m}=(n+1) {\cal H}_0\,   \phi_{n+1,m-1}^{\alpha_1\ldots\alpha_n\beta_1\ldots \beta_m}\; ,
    \qquad n,m=1,2,...\; ,
\end{eqnarray}
and
\begin{eqnarray}
\label{phi-n,1=0} &&
 \phi_{n,1}^{\alpha_1\ldots\alpha_n\beta_1}=0\; , \qquad
  \phi_{1,m}^{\alpha_1\beta_1\ldots \beta_m}=0\; .
    \qquad
\end{eqnarray}
Eqs. (\ref{phi-n,1=0}) and (\ref{phi-1+1=phi+1-1}) result in
\begin{eqnarray}
\label{phi-n,2l+1=0} &&
 \phi_{n,2l+1}^{\alpha_1\ldots\alpha_n\beta_1}=0\; , \qquad
  \phi_{2l+1,m}^{\alpha_1\beta_1\ldots \beta_m}=0\; ,
    \qquad
\end{eqnarray}
so that the decomposition in (\ref{Y=an}) has non vanishing coefficients only for even numbers $n$ and $m$. These are related by
\begin{eqnarray}
\label{phi-=phi+} &&
(2l+2)  \phi_{2k,2l+2}^{\alpha_1\ldots\alpha_{2k+2l+2}}=(2k+2) {\cal H}_0\,   \phi_{2k+2,2l}^{\alpha_1\ldots\alpha_{2k+2l+2}}\;,
\end{eqnarray}
which allows to express all the coefficients in terms of the ones for the monomial of $v_\alpha^-$ only,
\begin{eqnarray}
\label{phi2k2l=phi2k+2l-0} &&
 \phi_{2k,2l}^{\alpha_1\ldots\alpha_{2k+2l}}=\frac {(2k)!!}{(2l)!!} ({\cal H}_0)^l\,   \phi_{2k+2l,0}^{\alpha_1\ldots\alpha_{2k+2l}}\; .
    \qquad
\end{eqnarray}
  Thus the nontrivial solution the equations of our bosonic system is given by the series
\begin{eqnarray}
\label{Y=sol} &&
  \Upsilon = \sum\limits_{k=0}^\infty  ( \rho^\#)^{k} \phi_{2k,0}^{\alpha_1\ldots\alpha_{2k}}
\sum\limits_{m=0}^\infty  ( \rho^\#)^{-2l} \, \frac {(2(k-l))!!}{(2l)!!}\, v_{(\alpha_1}^+ \ldots v_{\alpha_{2l}}^+v_{\alpha_{2l+1}}^- \ldots v_{\beta_{2k})}^-\;   {\cal H}_0^l\; .
    \qquad
\end{eqnarray}

Thus we have seen that the bosonic limit of our 3D Matrix model superfield theory  gives a sensible Matrix model field theory  which has some nontrivial solutions.

\section{Conclusion}

In this paper we have constructed generalized Hamiltonian mechanics of the supersymmetric  non-Abelian multiwave system (nAmW) \cite{Bandos:2013swa} (which  is a simple
3D counterpart of the multiple M0-brane system \cite{Bandos:2012jz,Bandos:2013uoa}), and also performed its quantization by using the conversion and  generalized Gupta--Bleuler methods.

The set of dynamical variables of the nAmW system can be split on the set of center of energy variables and matrix variables describing the relative motion of $N$ nAmW constitutients. The former  is  same as the set of variables of  spinor moving frame formulation of massless superparticle; it includes, besides 3-vector bosonic $x^\mu$ and two component  fermionic spinor coordinates functions $\theta^\alpha$, also the spinor moving frame variables, the pair of bosonic spinors $v_\alpha^\mp$ constrained by (\ref{v-v+=1}). The relative motion variable sector is described by the set of Hermitian $N\times N$ traceless matrix fields, bosonic ${\bb A}_{\tau i}{}^j$, ${\bb X}_i{}^j$, ${\bb P}_i{}^j$ and fermionic
$\Psi_i{}^j$,  which can be obtained by dimensional reduction of 3D ${\cal N}=1$ SYM model down to d=1.

As the nAmW action \cite{Bandos:2013swa}  (in its simpler form) contains constrained spinor moving frame variables, it was convenient to introduce the covariant momenta, classical counterparts of covariant harmonic derivatives dual to the Cartan forms of the 3-dimensional Lorentz group. The canonical momentum for the fermionic matrix variable $\Psi_i{}^j$ is proportional to itself so that, after introducing an appropriate Dirac-Poisson brackets,
this field does not 'anticommute' with itself.
Using these ingredients and following the method by Dirac \cite{Dirac:1963}, we have found the complete set of constraints of the nAmW dynamical system, including two component spinor fermionic constraint generalizing the one of the massless superparticle mechanics, and split them on the sets of first and second class constraints.

We have found the Dirac  brackets allowing to threat the second class constraints as strong relations (in the sense of Dirac \cite{Dirac:1963}), but have chosen another way of dealing with second class constraints in preparing for the quantization. To this end we used the generalized Gupta-Bleuler method and conversion method which allows to pass to a gauge equivalent description of the system by generalized Hamiltonian mechanics with the first class constraints only.  In quantization the quantum counterparts of these can be imposed as equations on the quantum state vector of the system.

The quantization on the 3D nAmW system results in a system of equations for a state vector  superfield on superspace some of the coordinates of which are collected in traceless $N\times N$ matrices ${\bb X}_i{}^j$ (or ${\bb P}_i{}^j$), where $N$ can be interpreted as a number of constituents of our multi-wave system. These Matrix model field equations contain a fermionic matrix operators $\hat{\Psi}_i{}^j$
(quantum version of $\hat{\Psi}_i{}^j$) which does not anticommute with itself; its action  on the state vector superfield has to be defined.

As the algebra of these fermionic matrix operators $\hat{\Psi}_i{}^j$ is isomorphic to Clifford algebra with ($N^2-1$) generators, we have discussed briefly a Clifford superfield representation for the state vector, and in the case of $N=2$ elaborated this in a bit more detail.  
Actually this was a Clifford superfield depending on four ($N^2$) Clifford algebra valued variables  as we have also used Clifford superfield representation for the scalar fermionic operator  $\hat{\xi}$ corresponding to the additional conversion degree of freedom which has been introduced  to convert the second class fermionic constraint of our system into the first class constraint. (As in the case of single superparticle model, originally the fermionic spinor constraint of the nAmW system is the mixture of one  first class and one second class constraints).

We notice that the Clifford superfield representation will not be necessary 
in the case of multidimensional, D>3  nAmW systems. Indeed, already in D=4 the nAmW system (the action for which still is to be constructed) will include  two   complex conjugate counterparts of our Hermitian matrix operator $\hat{\Psi}_i{}^j$ and also two  component $\hat{\xi}$; this will allow us to split the set of these variables on the counterparts of creation and annihilation operators (complexified coordinates and momenta) and use the holomorphic (coordinate) representation to specify completely the set of equations of the 4D Matrix model field theory.

As a check of consistency, we have shown that the purely bosonic limit of our Matrix model field theory equations, which can be obtained by quantization of the purely bosonic limit of the 3D nAmW system, has a nontrivial solution. It would be interesting to find solutions of the complete supersymmetric system of equations.

The next problem for the future study is to construct the action, to develop generalized Hamiltonian approach for the above mentioned  D=4 cousin of our D=3 nAmW system, and to quantize it.
As we have already noticed, in this case it will be possible to find a simpler realization for the fermionic matrix operators, so that a more convenient description of D=4 Matrix model field theory will be available.
Our present study provides a 'proof of concept' basis for such a quantization of the 4D nAmW system.

Then an  interesting subject for future work will be the  quantization of the 11D mM0 system of \cite{Bandos:2012jz,Bandos:2013uoa}. While the quantization of massless 11D superparticle gives linearized 11D supergravity \cite{Green:1999by,Bandos:2007mi,Bandos:2007wm}, the quantization of mM0 system  should result in a prototype of an 11D free field theory on  (super)space with matrix coordinates which might be suggestive in search for exotic degrees of freedom of the hypothetical underlying M-theory.


\acknowledgments
{ The work by I.B. was supported in part by the
Spanish Ministry of Economy, Industry and Competitiveness  grant FPA 2015-66793-P, partially financed with FEDER/ERDF (European Regional Development Fund of the European
Union), by the Basque Government Grant IT-979-16, and the Basque Country University program UFI 11/55.
The work by M.S. was supported in part by CIIC 28/2018 and CONACyT program  ``Estancias sab\'aticas en el extranjero'',   grant 31065. }

\bigskip

\renewcommand{\theequation}{A.\arabic{equation}}

\begin{appendix}

\section{Quantization of massless particle in spinor moving frame formulation   and solution of the Klein--Gordon equation}
\label{3D}
\setcounter{equation}{0}

In this appendix we discuss the quantization of massless D=3 particle in spinor moving frame formulation and show how  the solution of the Klein-Gordon equation is reproduced in this procedure.

The massless particle action can be chosen to be
$$
S_{0}= \int\limits_{W^1}\rho^\# dx^a u_a^{=} = \int\limits_{W^1}d\tau \rho^\# \dot{x}{}^{\alpha\beta} v_\alpha^{-}v_\beta^-  \; , \qquad
$$
where $u_a^{=} $ is defined in (\ref{u--=3D}), $x^{\alpha\beta}= x^a\tilde{\gamma}_a^{\alpha\beta}$ and the spinor $v_\alpha^-$ is nonvanishing. It is convenient to consider it to be a column of spinor moving frame matrix the complementary element of which, $v_\alpha^+$, obeys  (\ref{v-v+=1}).

Calculating the canonical momenta (\ref{Pa=dL-ddx}) and covariant momenta (\ref{vPv=fbfd})--(\ref{d--:=}), we find the following set of
{\it primary constraints} (in terminology of  \cite{Dirac:1963})
\begin{eqnarray}
\label{Phia=Pa-=00}
{\Phi}_a &:= & P_a -  {\rho}^\# u_a^= \approx 0 \; , \qquad
 \\ \label{d--=00}
 && {\mathbf d}^{=}\approx 0 \; , \qquad
 \\ \label{d++=00}
 && {\mathbf d}^{\#}\approx 0 \; , \qquad
 \\ \label{td0=d0-=00}
 \tilde{{\mathbf d}}{}^{(0)}&:=& {\mathbf d}^{(0)} -2 {\rho^{\#}} P_{_{\rho^{\#}}} \approx 0 \; , \qquad
  \\ \label{Prho=00}
 && P_{_{\rho^{\#}}}\approx 0 \; . \qquad
\end{eqnarray}
The first constraint can be split onto
\begin{eqnarray}
\label{Phi++=0}
&& \Phi^\# := \Phi_a u^{a\#}=  P_a u^{a\#} - 2 {\rho}^\# \approx  0 \; , \qquad \\
\label{Phi--=0}
&& \Phi^=:= \Phi_a u^{a=}:= P_a u^{a=}  \approx 0 \;
   \;, \qquad \\
\label{PhiI=0}
&& \Phi^\perp := \Phi_a u^{a\perp}\approx  0 \; .
\qquad\end{eqnarray}
Introducing the Poisson brackets (\ref{PB=Px}) and (\ref{bfdv=}), one observes that (\ref{Phi++=0}) and (\ref{Prho=00}) form a pair of second class constraints. The same applies to   (\ref{PhiI=0}) and (\ref{d++=00}).

As in the main text, we use the generalized Gupta-Bleuler approach to deal with such pairs of second class constraints. Namely we just relax one of the conjugate constraints  thus promoting the second element of the pair to be the first class constraints. Similarly as it was done in the more complicated nAmW case, we relax the second class constraints  (\ref{Prho=00}) and  (\ref{d++=00}) thus arriving at the dynamical system with the first class constraints
(\ref{Phia=Pa-=00}), (\ref{d--=00}) and (\ref{td0=d0-=00}).

Quantizing the system in the coordinate representation (see the relevant equations in (\ref{coord-repP}), (\ref{coord-repX}), (\ref{D=:=}))
and imposing the constraints on the wavefunction we arrive at the following set of equations\footnote{As in the more complicated case of nAmW, the canonical Hamiltonian of the massless particle is weakly equal to zero and the Schr\"odinger equation implies just independence of the state vector on the proper time variable.}
\begin{eqnarray}
\label{partXi=0}
&& \left( \partial_a -  i{\rho}^\# u_a^=  \right) \Xi = 0 \; , \qquad \\
\label{D--Xi=0} &&
 D^=  \Xi = 0 \; , \qquad \\
\label{D0Xi=0}
&& \left( D^{(0)}
+2 {\rho^{\#}} \frac \partial {\partial \rho^\#} \right) \Xi = 0 \; . \qquad \end{eqnarray}

Eq. (\ref{D--Xi=0}) implies independence of the state vector $\Xi$ on the complementary spinor moving frame variable $v_\alpha^+$,
$\Xi=\Xi(x^a,v^-,\rho^\#)$. Eq. (\ref{partXi=0}) fixes the dependence on the spacetime coordinate to be
\begin{eqnarray}
\label{Xi=expY}
\Xi (x^a,v^-,\rho^\#)= e^{i\rho^\# x^a u_a^= } Y (v^-,\rho^\#)
\; . \qquad \end{eqnarray}
Then Eq. (\ref{D0Xi=0}) results in
\begin{eqnarray}
\label{D0Y=0}
&& \left( D^{(0)}
+2 {\rho^{\#}} \frac \partial {\partial \rho^\#} \right)  Y (v^-,\rho^\#) = 0 \; , \qquad \end{eqnarray}
which implies that $Y$ actually depend on a particular product of $\sqrt{\rho^\#}$ and $v^-$, on the nonvanishing bosonic spinor $\lambda_\alpha:=\sqrt{\rho^\#} v^-_\alpha$, only:
\begin{eqnarray}
\label{Y=Yl}
&&
Y=Y(\lambda_\alpha)\; , \qquad \lambda_\alpha:=\sqrt{\rho^\#} v^-_\alpha\; . \qquad \end{eqnarray}
Then, as far as  $\rho^\#u_a^= = \lambda \tilde{\gamma}_a\lambda$ (see (\ref{u--=3D})), our state vector is represented by
\begin{eqnarray}
\label{Xi=expYl}
\Xi = \Xi (x^a, \lambda) = e^{i\lambda\tilde{\gamma}_a\lambda\, x^a} Y (\lambda)
\; .
\qquad
\end{eqnarray}
As  $\lambda$ is a square root of the particle momentum, one observes that our state vector is actually depend on both coordinate and momentum degrees of freedom. It is a counterpart of basic plane wave solution of the standard formalism, which should be integrated over momentum to arrive at the field depending on the coordinates only.

Similarly in our case the representation  by functions depending on coordinates can be reached by performing an integration over
bosonic spinors $\lambda_\alpha$. In such a way we arrive at
\begin{eqnarray}
\label{phi-x=}
\phi (x) = \int d^2 \lambda \,  \Xi (x^a, \lambda) =  \int d^2 \lambda \,  e^{i\lambda\tilde{\gamma}_a\lambda\, x^a} Y (\lambda)\; .
\qquad
\end{eqnarray}
It is not difficult to check that the scalar field (\ref{phi-x=}) obeys the Klein-Gordon equation
$$
\Box \phi (x) := \partial_a\partial^a \phi (x) = 0\; .
$$
Thus we have shown that the quantization of massless particle by the method applied in the main text to nAmW system results (as it should) in a theory of free scalar field obeying the massless Klein-Gordon equation.

\end{appendix}

\end{document}